\documentclass[final]{svjour3}
\usepackage{graphicx}
\usepackage{rotating}
\usepackage{amssymb}
\usepackage{mathptmx}
\usepackage[square,numbers]{natbib}
\makeatletter
\journalname{}

\begin{document}

\title{Phonon and charge signals from IR and X excitation in the SELENDIS Ge cryogenic detector}

\author{H. Lattaud \and Q. Arnaud \and J. Billard \and  J. Colas \and L. Dumoulin \and M. De Jésus \and  A. Juillard \and  J. Gascon \and  S. Marnieros \and  C. Oriol.}

\institute{H. Lattaud \and Q. Arnaud \and J. Billard \and  J. Colas \and M. De Jésus \and  A. Juillard \and  J. Gascon \\ Univ. Lyon, Univ. Lyon 1, CNRS/IN2P3, IP2I Lyon, F-69622, Villeurbanne, France\\
\email{lattaud@ipnl.in2p3.fr} \\
\\
L. Dumoulin \and S. Marnieros \and  C. Oriol \\
Université Paris-Saclay, CNRS/IN2P3, IJCLab, 91405 Orsay, France.
}

\maketitle

\begin{abstract}

The aim of the SELENDIS project within the EDELWEISS collaboration is to observe  single $e^- h^+$ pairs in  lightweight (3.3 g) cryogenic germanium bolometers  
with charge and phonon readout at biases up to $\sim 100$ V. These devices are ideal to characterize in detail the mechanism of charge creation and collection 
in cryogenic germanium detectors. Electron-hole pairs are produced in the bulk of the detector either by the injection of pulsed IR laser or by neutron activation of germanium inducing the K, L and M lines from  $^{71}$Ge electron capture decays. Low-energy laser pulses are also used to probe the single $e^- h^+$ pair sensitivity of Ge bolometers.
Preliminary results are used to compare these two modes of charge creation, an important step toward a detailed characterization of Ge bolometers for their use in sub-MeV Dark Matter (DM) searches.


\end{abstract}

\section{Introduction}
The EDELWEISS collaboration is working on the direct detection of DM since the 90s using bolometric semiconductor detectors. Originally, the aim of those detectors was to probe DM masses ranging from $\sim 40$ GeV to $1$ TeV \cite{lastEdel}. Recent developments and experimental constraints \cite{CDMS,edelweiss} have shown that semiconductor bolometers are well suited to probe lower mass DM particle candidates. 
In order to be competitive, detectors which energy resolution reach sensitivity to single $e^- h^+$ pair signals \cite{threshold}, will be able to explore models extending from the eV to the MeV mass range \cite{mod1,mod2,mod3}.  In this context, the SELENDIS program  develops prototypes of lightweight Ge bolometers able to sustain high biases to reach energy resolutions better than half the $e^- h^+$ pair creating energy in Ge ($3$ eV). In the course of this development, we have studied the response of Ge detector to low energy IR-photons. 
In this paper, the Neganov-Trofimov-Luke (NTL) effect \cite{luke} is used to extract the quantum efficiency $\eta$ of $e^- h^+$ pair creation \cite{Domangue2009} in SELENDIS detector for IR-photon of various wavelengths close to the band gap of Ge ($1310$, $1550$ and $1650$ nm). The wavelength yielding the highest efficiency is selected to probe the single $e^- h^+$ pair sensitivity of  SELENDIS detector prototypes, injecting laser pulses corresponding to the creation of a few pairs.
This paper is organized as follows. The experimental procedure is described in Sec.~\ref{sec1}.  The quantum efficiency  for $e^- h^+$ pairs creation is presented and discussed in Sec.~\ref{sec2}.  The first results of single $e^- h^+$ pair sensitivity of the SEL40 detector prototype are presented in Sec.~\ref{sec3}. Finally, we draw the conclusion in Sec.~\ref{sec4}.

\section{Experimental setup}\label{sec1}
The two detector prototypes (SEL30 and SEL40) are cylindrical high-purity Ge crystals of 14 mm diameter and 4 mm height, corresponding to a mass of $3.3$ g. A hydrogenated amorphous Ge layer (\textit{a}Ge-H, thickness of $30$ nm ) is deposited onto the surfaces of SEL30 while none are on SEL40. Grid patterned aluminum electrodes (thickness of $150$ nm, width of $20\, \mu $m, pitch of $300\, \mu $m) are lithographed on both top and bottom surfaces. There are no electrodes on the lateral surface to avoid leakage currents when operated at high biases. A  $2\times 2$  mm$^2$ Ge neutron-transmutation-doped (NTD) heat sensor \cite{NTDpaper} is glued on the top surface and thermally coupled  to the copper housing through gold wire bonds. 
The signal readout consists of two ionization channels and one heat channel.
Those detectors are integrated in the dry dilution cryostat at \textit{IP2I Lyon}, a Hexadry-200 cryostat from Cryoconcept \cite{cryoconcept}. The cold and warm electronics are those described in Ref.~\cite{electroniclyon}. This setup, which is the same as the one described in Ref.~\cite{red30} allows to operate the detectors at cryogenic temperatures down to 14 mK. Three $10$ mW laser diodes (Tab.~\ref{tab1})  from Aerodiode \cite{aerodiode} operated outside  the cryostat are used to illuminate the detectors. The optical signal is guided through optical fibers thermalized at each stage of the cryostat and directly mounted on the detector's copper housing. The optical power can be modulated using a set of attenuators and the pulse width, allowing to vary the energy deposited per pulse. The data processing and analysis are performed using the algorithm described in Ref.~\cite{MPS} and the ROOT framework \cite{rootframework}.

\begin{figure}[h!]
\begin{center}
\includegraphics[width=0.9\linewidth]{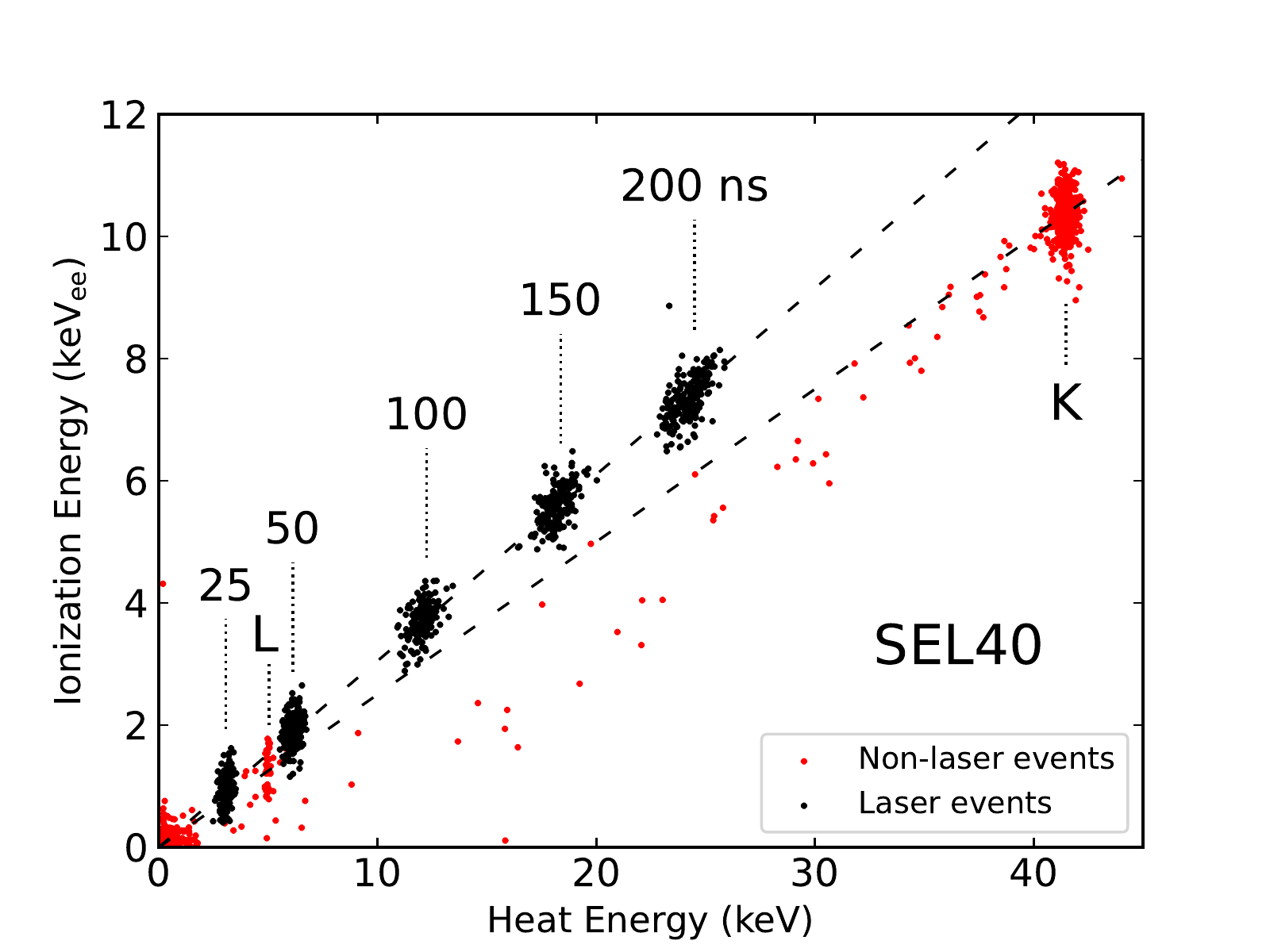}
\caption{2D Energy deposit (ionization and heat) for the K and L lines (red dots) and laser energy deposit (black dots) with pulse width scan, step from $25$ to $200$ ns: (Color figure online.)}\label{laserresponses}
\end{center}
\label{fig1}
\end{figure}

Prior to the installation in the cryostat, the detectors have been activated during 4 days using a strong AmBe neutron source. This activation produces the isotope  $^{71}$Ge  which decays with a half-life of $11$ days by electron capture in the K, L, M shells, producing X-rays with the total energy of $10.37$, $1.3$ and $0.16$ keV.   Such lines are absorbed in the bulk of the detector, probing the detector response and allowing a precise low-energy calibration.
The 2D energy spectrum  (heat and ionization) is presented in Fig.~\ref{laserresponses}. Those data have been recorded applying a $9$ V bias on SEL40,  with an observed baseline energy resolution of $26.3$ eV (RMS).  A linear increase of the laser diode energy deposit as a function of the pulse width is observed, as well as a non-linearity of $4.9\%$ in the energy scale between high and low energy.

\section{Quantum efficiency of $e^- h^+$ pair production by IR radiation }\label{sec2}

For the extraction of the quantum efficiency, the data taking is done after detector reset (all electrodes grounded). The reset procedure consists of an irradiation with a $^{137}$Cs source in order to neutralize accumulated space charges. The data has  been recorded in one afternoon, switching between the different laser diodes. No degradation of the charge collection of the K-line has been observed. The pulse width is fixed to an arbitrary value to produce an energy deposit smaller than the one created by the de-excitation of the K-line, and then the bias applied to the electrode is varied between $1$ and $12$ V.

The analysis is based on the study of the energy dissipated by the drift of the moving charges within the electric field, the NTL effect \cite{lukeneganov1,lukeneganov2}. 
An absorbed photon will deposit its energy $h\nu$ via the production of excitons or phonons \cite{exciton}, with the possibility of creating some $e^- h^+ $ pairs. All that energy is eventually converted into thermal phonons once all charges either recombine or are collected. A burst of $N$ photon will create on average a total number of pairs $N_{pairs} = \sum _{i = 0}^{N} i\times N_{i}$, where $N_i$ is the average number of photons producing a number of pairs $i$.  At the wavelengths considered here, a single photon cannot produce more than one $e^- h^+ $ pair: $N=N_0+N_1$ and $N_{pairs} = N_1$.  With the application of the bias $V$, the heat energy $E_{phonon}$ will be further increased by the NTL heating:  
\begin{equation}\label{eq1bis}
    E_{phonon} (V) = h\nu \times N_0 +  h\nu \times N_{1}  + |V| \times N_{1}    
\end{equation}
By defining the quantum efficiency of pair creation $\eta$ as:
\begin{equation}\label{eq2}
	\eta = \frac{N_{1}}{N_0+N_{1}}
\end{equation}
Eq.~\ref{eq1bis} can be written: 
\begin{equation}\label{eq1}
	E_{phonon} (V) = h\nu \times (N_0 + N_{1}) \times ( 1 + \eta \frac{|V|}{h \nu}  )
\end{equation}

\begin{figure}[h]
\begin{center}
\includegraphics[width=0.5\linewidth]{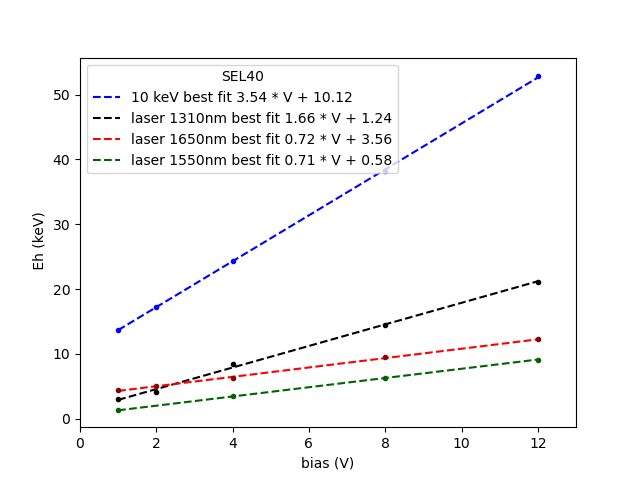}
\includegraphics[width=0.49\linewidth]{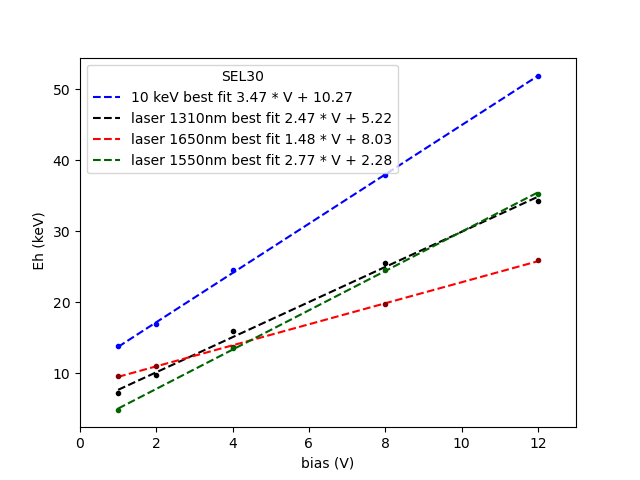}
\includegraphics[width=0.9\linewidth]{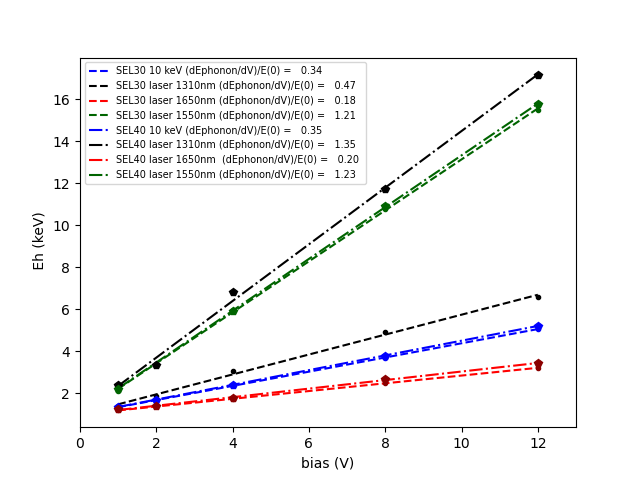}
\caption{{\it Top left, Top right }  heat signal (in keV)  as a function of the applied bias (in V) for SEL30 and SEL40, respectively. {\it Bottom} heat signal normalized to the deposited energy at $0$ V as a function of the applied bias: (Color figure online.)}\label{fig2}
\end{center}
\end{figure}

The results (Fig.~\ref{fig2}) for the three different wavelengths and the K line show that the heat energy $E_{phonon}$ depends linearly on the applied bias as expected from Eq.~\ref{eq1}. A  simple relationship for each wavelength can be extracted by normalizing the data to the extrapolated energy at $V=0$. This allows to extract the quantum efficiency $\eta$ as follows:
\begin{equation}\label{eq3}
	\eta= \frac{h \nu \times \frac{dE_{phonon}}{dV} }{E_{phonon}(0) }
\end{equation}

Each laser diode is characterized by  a single $\eta$ parameter, determined using the slopes extracted from a linear fit to the data, as presented in Tab.~\ref{tab1} which combines the measurements obtained by the two detectors.

\begin{table}[h!]

\begin{center}
 \begin{tabular}{|c | c c c c|} 
  \hline
 Emission  wavelength (nm) & 1310 & 1550 & 1650 & K line \\  

 \hline
 Photon energy (eV)& 0.95 & 0.8 & 0.75 & 10.37\\ 
 \hline
 Absorption length in Ge (mm) \cite{absGe1,absGe2} & 1e-4 & 2.1 &  17 & \\ 
 \hline 
  Combined $\eta$ & $0.855 \pm 0.415 $ & $0.98 \pm 0.05$ & $0.146 \pm 0.01$ & $1.03 \pm  0.05$ \\ 
 \hline
\end{tabular}
\end{center}
\caption{Laser characteristics and combined measured quantum efficiency from SEL30 and SEL40.}\label{tab1}
\end{table}

For the K line, the value of $h \nu$ can be replaced by the pair creation energy  $\epsilon _{\gamma} = 3$ eV, since the emitted photons are produced homogeneously within the detector and carry an energy well above the Ge direct gap. In that case, a value of $\eta = 1 $  is a check that the energy calibration is correct and that energy scale non-linearity effect is negligible.
For the $1310 $ nm laser, the efficiency is around $ 85\%$ with a large uncertainty that comes mainly from the discrepancies observed in both detectors  ($47\%$ measured in SEL30 and $135\%$ measured in SEL40). These discrepancies can be interpreted using the characteristics of the photons. The energy being higher than the direct gap in Ge ($0.77$ eV \cite{Domangue2009}), each photon can create an $e^- h^+$ pair but due to the short absorption length of such photon in Ge (Tab.~\ref{tab1}) those charges are created at the surface of the detector, where the charge collection is not optimal due to trapping at low bias and incomplete charge collection \cite{cdmssurface}. 
For the $1650$ nm laser, the efficiency is $14.6\%$. The energy is close but below the direct gap, and due to the large absorption length (Tab.~\ref{tab1}) the charges are created homogeneously within the detector volume. The photons can either create a $e^- h^+$ pair by absorption or, when the energy is not sufficient, produce short-lived  bound excitons which recombine by non-radiative processes and thus  contribute to the heat energy but without adding charge signals \cite{exciton}.
For the $1550$ nm laser, the efficiency is $98\%$. This wavelength  optimizes the compromise between two trends: the absorption length is large enough so that the interactions are in the bulk and  thus avoid the inefficiencies due to charge trapping, but at the same time the energy is large enough to have an important probability to create a pair.

\section{Probing single $e^- h^+$ pair sensitivity of SEL40}\label{sec3}

\begin{figure}[h]
\centering
\includegraphics[width=0.50\linewidth]{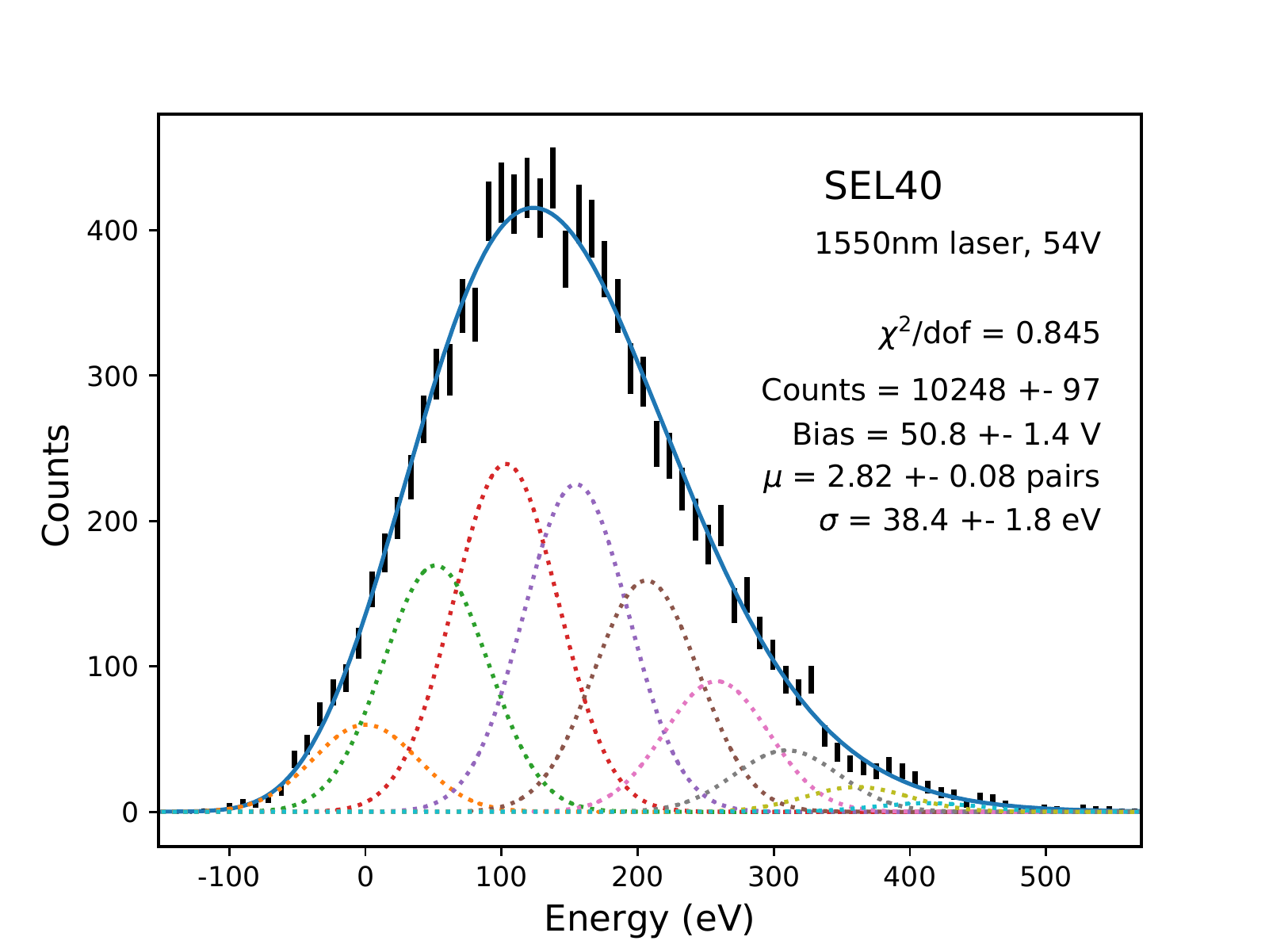}
\includegraphics[width=0.49\linewidth]{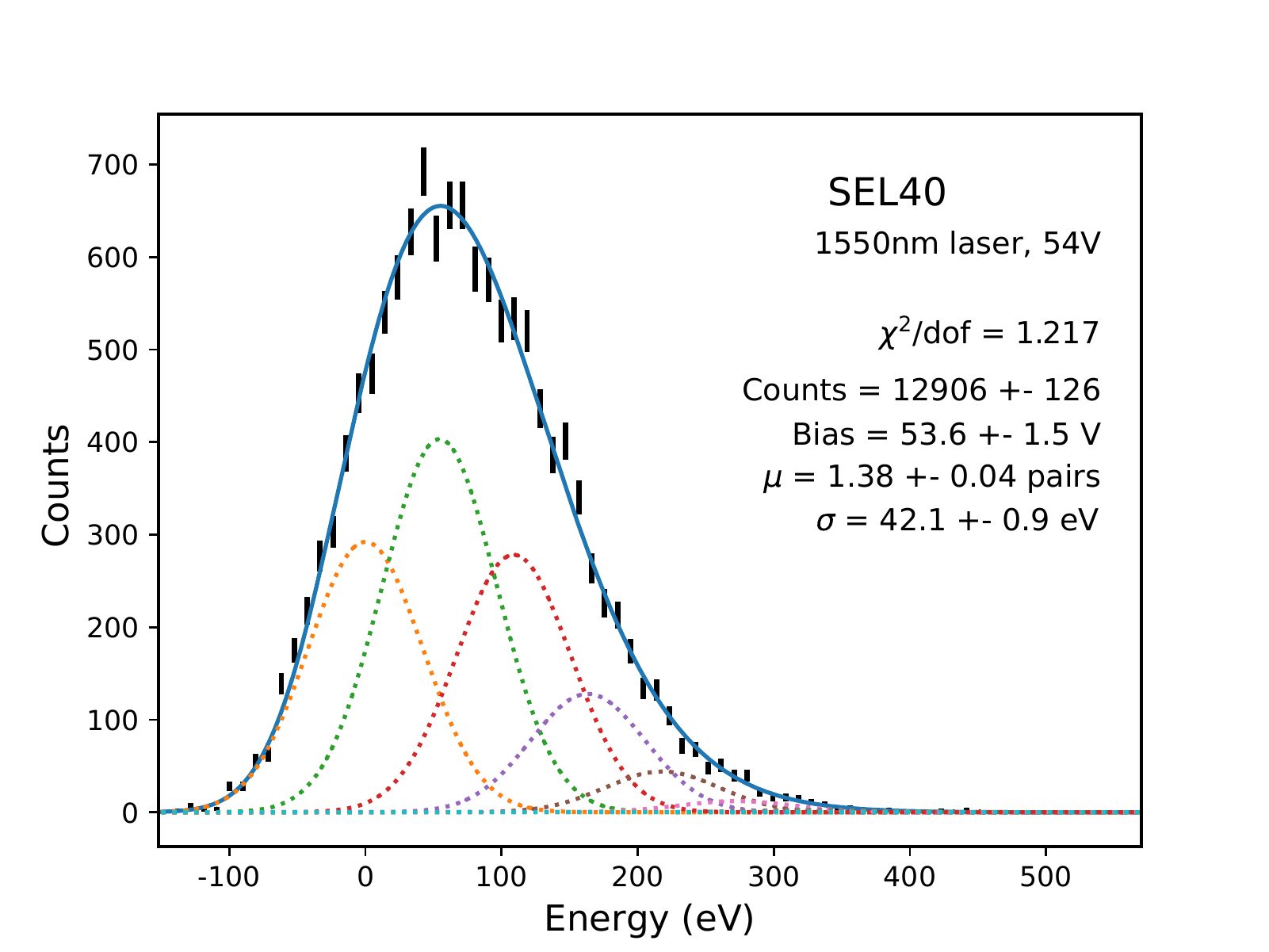}
\caption{{\it Left, right,} Heat energy spectra from two laser configurations with energy deposits $2.8$  and $1.4$ pairs, respectively. Black dots are the data, colored gaussians show the expected Poisson distribution for such energy smeared by detector resolution,  blue line shows the fit of the Poisson distribution smeared by a gaussian of $\sigma \approx 40 $ eV: (Color figure online.)}\label{fig3}
\end{figure}

The sensitivity of  SELENDIS detectors to few $e^- h^+$ pairs events could be probed so far with only one prototype, SEL40, as it achieved the best heat energy resolution.
Following the results of  Sec.~\ref{sec2}, we chose the $1550$ nm laser for this study, to benefit from homogenous charge carrier creation and high quantum efficiency.
Prior to the data taking, we applied the same reset procedure presented in Sec.~\ref{sec2} to reduce charge trapping. The biases are slowly increased until the highest bias that the detector can withstand without a significant addition of noise is reached. For SEL40, the bias is $54$ V. In these operating conditions, the heat signal is  boosted by a factor $19$ through NTL effect. 
In order to reduce data contamination by unwanted backgrounds and lower the energy threshold of the analysis, a pulse amplitude is fitted by imposing that it occurs in coincidence with the known time that it was injected by the laser diode. It was checked that the average pulse height distribution obtained at other times is centered on zero with dispersion corresponding to the heat baseline energy resolution, $\sigma _{phonon}$. Fig.~\ref{fig3} shows the phonon signal recorded in coincidence with laser pulses having an amplitude calculated to correspond to an average energy deposit $\mu$ of $2.8$ ({\it left}) or $1.4$ ({\it right}) pairs. The amount of attenuation and pulse width are extrapolated from the values of $N_{pair}$ obtained with different setting  condition from studies of  Sec.~\ref{sec2}.

The phonon energy is obtained from the K-line calibration, in keVee, converted in true total energy using the relation $E_{total}=E_{keVee} \times (1+ |V| / 3) $ to correct for the NTL effect. The production of $N_{pair}$ should thus result with a peak at $E_{total}=N_{pair} \times |V|$, with a width of $\sigma _{phonon}$. As photons are deposited at random depth across the detector, the distribution of $N_{pair}$ for a given $\mu$ value should obey the Poisson statistics. The data of Fig.~\ref{fig3} are thus adjusted with a Poisson distribution, where each $N_{pair}$ contribution is smeared by a gaussian of width $\sigma _{phonon}$, centered at the position $N_{pair} \times |V|$ \footnote{$\eta = 1$ is assumed for $1550$ nm photons}. To take into account possible non-linearity between the calibration at high energy and these low-energy signals, the global energy scale is left as a free parameter. The distribution are thus fit with only 4 parameters: $N_{count}$, $V$, $\mu$ and $\sigma$. As the data are triggered by the pulser system, no background term is included in the fit. The results are shown as blue lines in Fig.~3.
This simple model nevertheless gives $\chi ^2$ close to $1$, and the fitted $V$ and $\sigma$ values correspond   to the expected values within $\sim 2\sigma$, and the ratio of the $\mu$ values ($2.82$ and $1.38$) corresponds to the expected one as the pulse width was divided by a factor $2$ between the two dataset.
The resolution of $40$ eV (RMS) is not sufficient to visually resolve the individual $N_{pair}$ contributions at $54$ V but the shape of the Poisson distribution  convincingly shows the presence of the different  $N_{pair}$ contributions with the expected intensity.

Although not entirely resolved, the presence of single-pair events in the data is clearly established. This demonstrates that cryogenic detector with the NTL amplification can detect single-charge event, a fact that was alluded to but not demonstrated in Ref.~\cite{edelweiss}. In that work, the detector had a similar baseline energy resolution of $42$ eV (RMS), combined with a higher bias of $78$ V that lead to a resolution of $0.53$ pairs, instead of the value obtained here of $0.75$ pairs. However, the laser-triggered setup developed here was not available then. The next step of this study is to apply the same method to detectors with better energy resolution and higher biases such as the one in Ref.~\cite{edelweiss} and other future detectors Ref.~\cite{gasconltd}.

\section{Conclusion}\label{sec4}
The first results of the SELENDIS project have been presented. The Ge response to photon absorption with IR-laser pulses have been probed to extract the quantum efficiency of pair creation by such photons. The best suited wavelength regarding the quantum efficiency $\eta$ is the $1550$ nm reaching nearly $100\%$ of efficiency. This laser has been used to probe the sensitivity of a SELENDIS detector to single $e^- h^+$ pair by applying high biases. Whereas,  single peak discrimination was not achieved, the results show good compatibility to  the expected Poisson distribution of pairs. The next steps are to work on detector design to lower the resolution and reach higher biases. Those results are of major interest for the research on the design of the next generation of detectors for the EDELWEISS-SubGeV project \cite{gasconltd}.

\begin{acknowledgements}
We acknowledge the support from the EDELWEISS Collaboration, the Ricochet Collaboration,  LabEx Lyon Institute of Origins (ANR-10-LABX-0066), the European Research Council  (ERC)  under  the  European  Union’s  Horizon  2020  research  and  innovation  program under Grant Agreement ERC-StG-CENNS 803079 and Marie Sk\l{}odowska-Curie Grant Agreement  No. 83853.
\end{acknowledgements}

\pagebreak


\begin{thebibliography}{99}
\bibitem{lastEdel}
L. Hehn et al. (EDELWEISS Collaboration),  Eur. Phys. J. C (2016) 76: 548
https://doi.org/10.1140/epjc/s10052-016-4388-y

\bibitem{CDMS}
R. Agnese et al., Phys. Rev. Lett. 121, 051301, Erratum Phys. Rev. Lett. 122, 069901 (2019)

\bibitem{edelweiss}
Q. Arnaud et al. (EDELWEISS Collaboration), Phys. Rev. Lett. 125, 141301 

\bibitem{threshold}
M. Battaglieri et al., US Cosmic Visions: New Ideas in
Dark Matter 2017: Community Report, arXiv:1707.04591

\bibitem{mod1}
H. An, M. Pospelov, J. Pradler, and A. Ritz, Phys. Lett.
B 747, 331 (2015) arXiv:1412.8378

\bibitem{mod2}
Y. Hochberg, T. Lin, and K.M. Zurek, Phys. Rev. D 95,
023013 (2017) arXiv:1608.01994

\bibitem{mod3}
R. Essig et al., J. High Energ. Phys. 2016, 46 (2016)
arXiv:1509.01598



\bibitem{luke}
M. P. Chapellier et al., Physica B, 284-288 (2000) 2135-2136
\bibitem{NTDpaper}
Eugene E. Haller et al., Proc. SPIE 2198, Instrumentation in Astronomy VIII, (1 June 1994); https://doi.org/10.1117/12.176771

\bibitem{Domangue2009}
 J. Domange, A. Broniatowski, E.  Olivieri, M. Chapellier, L. Dumoulin, {\it AIP Conference Proceedings} \textbf{1185}, 1, 314-317, (2009), DOI:10.1063/1.3292341

\bibitem{cryoconcept}
http://cryoconcept.com/, (accessed on Oct, 2021)

\bibitem{electroniclyon}
E. Armengaud et al. (EDELWEISS Collaboration), JINST 12, P08010 (2017), arXiv:1706.01070.

\bibitem{red30}
Q. Arnaud et al. (EDELWEISS Collaboration), Phys. Rev. Lett. 125, 141301 

\bibitem{aerodiode}
https://www.aerodiode.com/, (accessed on May 18th, 2021)

\bibitem{MPS}
J. Colas and the RICOCHET collaboration, J. Low Temp. Phys. This Special Issue (2021)

\bibitem{rootframework}
R. Brun and F. Rademakers, Proceedings of AIHENP'96 Workshop, Lausanne, Sep. 1996,
Nucl. Inst. and Meth. in Phys. Res. A 389 (1997) 81-86.

\bibitem{absGe1}
S.M. Sze, Physics of Semiconductor Devices, Wiley,
1981.
\bibitem{absGe2}
G. G. Macfarlane et al., Phys. Rev. 108 (1957) 1377-
1383.

\bibitem{lukeneganov1}
B. Neganov and V. Trofimov, USSR patent No. 1037771, 1981; Otkrytializobreteniya 146 (1985) 215.

\bibitem{lukeneganov2}
P.N. Luke. Appl. Phys. 64 (1988) 6858.

\bibitem{cdmssurface}
Ponce, F., Page, W., Brink, P.L. et al. Modeling of Impact Ionization and Charge Trapping in SuperCDMS HVeV Detectors. J Low Temp Phys 199, 598–605 (2020). https://doi.org/10.1007/s10909-020-02349-x

\bibitem{exciton}
J. C. Culbertson, R. M. Westervelt and E. E. Haller,
Phys. Rev. B 34 (1986) 6980-6986


\bibitem{gasconltd}
J. Gascon et al., J. Low Temp. Phys. This Special Issue (2021)

\end{thebibliography}
\end{document}